# Continue the Analogy of Physics and Economics.
## Self-induced Transparency Mechanism as an Invisible Hand of Market

A.V. Samokish, V. E. Egorushkin

1. **Introduction.**

Economics is a high-developed branch of economic science which successfully describes the economic laws with a physical methodology [1–27]. The methods of Gauge Field Theory [1], Electrodynamics [2,3], Statistic and Thermodynamics [4–6], Plasticity [7], Mechanics and Quantum Mechanics [8–10] have been used to explain many economical phenomena. It includes equilibrium and dynamic of market [10], financial instability [5], Phillips curve, recession, Black-Scholes equations [2], demand and supply growth and economic cycles [4].

It is important to outline that these two branches – economics and physics – have very deep connections. Inflation in economic market, electricity or plasticity in physics, when appeared, break initial dynamic equilibrium, so that old dynamics isn't capable to reflect new conditions. These conditions interrupt the motion of money flow as well as motion of physical particles or displacements in the phase space of physical or economic states. The phase space in economics, as in physics, represents an abstract space with orthogonal coordinates necessary for defining the state of the economy.

For various economic phenomena, coordinates of the phase space can be assigned differently [10, 11]. In the case of financial markets, the phase space of the initial equilibrium money dynamic possesses characteristics of quantum mechanics space [12]. The emergence of strong inflation or structural changes in the market can't be compensated neither by introduction of the real money market [13] nor with the changes in the initial equilibrium. General market behavior under such conditions is explained by introducing the "invisible hand of the market" – competition, profit, choice, etc. [14].

The invisible hand can regulate economic activity, prices in supply-demand processes, resource allocation to their most efficient uses and development of new technologies. However, the invisible hand is a theoretical construct, and its working mechanism is still a mystery [15]. Nonetheless, the restoration of a dynamic framework accounting for the altered structure of state spaces is made possible by introducing curved or fiber spaces. In this structure, the movement of particles, money and other market mechanisms is defined by parallel transport [16].

In our previous work [7], we demonstrated an analogy between the gauge theory of plasticity and economics, and argued that this analogy is far from accidental. The similarity in the behavior of plastic flow and economic choice (self-interest) in an inflationary market is due to the fact that, in both cases, these processes are described by a gauge field that restores equilibrium disturbed by defects in physics and by inflation in economics. The emergence of this field is driven by local changes in the Berry phase [17], which, in the economic context, represents the complete profit of the market — the fundamental objective of economic relations. The forces generated by this field — competition and local profit — restore equilibrium under inflationary conditions. In a financial market without inflation, all quantities that determine equilibrium change uniformly [9,10,18]. In a market with inflation, the increase in the money supply (DM) is the source of the field that restores the equilibrium disrupted by inflation [10,18]. This field mediates the interaction between sources. The potential of the field is Choice (self-interest), while its intensity is given by Competition and Profit — the forces with which the field acts on both resting (unchanged in economic phase space) and moving increments of DM. It is precisely through these forces that a new, non-uniform equilibrium is achieved in an inflationary market, distinct from the previous, uniform equilibrium.

Because the market is non-uniform, the field and its characteristics are local. In order not to destroy the emerging equilibrium, the field is forced to propagate in the form of waves — in this case, in economic phase space. The equations describing the spatial behavior of the field in this space are nonlinear wave equations defined on local regions. As a consequence, the field characteristics – Choice ($\vec{Ch}$), Competition ($\vec{C}$), and Profit ($\vec{P}$) flows are also local nonlinear wave processes. The solutions of the equations for $\vec{Ch}, \vec{C}, \vec{P}$ describe the evolution of the market. The spatio-temporal evolution of choice is particularly informative: the choice curve captures the overall state of the economy and all motivational structures involved in achieving its main objective. Constructing and analyzing this curve makes it possible to shed light on possible scenarios of economic development and to turn the invisible hand of the market into a visible one.

In this paper we begin from the moment when the above fields arise – that is, from the point where the topological features of the economic phase space become essential, something that had not been addressed in [17].

## 2. Berry Phase, Profit and Choice

The phase space in which the evolution of the economic states of the initial equilibrium market takes place is assumed to be unitary, as is the evolution itself [1–3]. The distribution properties of money (securities holdings and cash) and the representation of the total money supply M in the form

$$M = \sum_k C_k m_k \qquad (1)$$

together with all possible realizations of the market state, imply that phase space (PS) is a vector, unitary, linear, complex orthogonal space endowed with an admissible norm [1,2,4]. In this setting, the evolution of states can be written as [1–4]:

$$\frac{\partial \ln \vec{M}}{\partial t} = -iA \qquad (2)$$

where A is the level of economic activity of the market with the dimension of a frequency of market events, defined relative to the growth rate of the money supply DM. Equation (2) can be equivalently rewritten as:

$$\frac{\partial \vec{M}}{\partial t} = -iA\vec{M} \qquad (3)$$

Taking the scalar product of (3) with $\vec{M}$, we obtain:

$$A = i\left(\vec{M}\frac{\partial \vec{M}}{\partial t}\right) \qquad (4)$$

which is a definition of market activity expressed through the dynamics of $\vec{M}$ in PS.

To analyze the topological properties of PS, we substitute (1) into (3) and obtain:

$$\sum_k (\dot{C}_k m_k + C_k \dot{m}_k - iAC_k m_k) = 0 \tag{5}$$

Here the dot denotes a derivative with respect to time.

Calculating the dot product of (5) with $m_n$ we obtain:

$$\dot{C}_k \delta_{kn} + C_k (m_n \dot{m}_k) + iAC_k \delta_{kn} = 0 \tag{6}$$

Solving (6), we get:

$$C_n = e^{-\int_0^t [(m_n \dot{m}_k) + iA] dt'}, \quad n = k \tag{7.1}$$

$$(m_n, \dot{m}_k) = 0, n \neq k \tag{7.2}$$

In (7.1) the integral $\int_0^t A \, dt'$ defines the dynamical phase of the moving money supply $M_k$. If A doesn't depend on time $A = const$, the market is in a stationary state, without any shocks. The value of $\int_0^t [(m_n \dot{m}_k) + iA] \, dt'$ – is the topological phase of the vector $\vec{m}$, arising from changes in the character of the motion of $\vec{m}$ in the phase space. The condition (7.2) – $(m_n, \dot{m}_k)$ is precisely the condition of parallel transport [16] of the vector $\vec{m}$ in PS, which replaces the original dynamics of $\vec{m}$. Under these new conditions the explicit dependence on time ($t$) is transferred into the dependence on the phase-space coordinates $X(t)$.

Here it is appropriate to note that the quantity $(m_n, \dot{m}_k)$ is purely imaginary and does not represent the real (material) economic world but provide a picture of an imaginary informational world. In this ideal world Choice (self-interest) determines the flows of prices, resources, goods, competition, profit and information governed by the laws of the market – its **invisible hand.**

Explicitly, the phase of $\vec{m}_n$ can be represented in the following form:

$$\vec{m}_n(t) = e^{i\Theta(X(t))} \tilde{m}_n(X(t)) \tag{8}$$

where $\tilde{m}_n$— is the certain set of the money supply (DM).

Substituting (8) into (7.2), we get:

$$\dot{\Theta}(X(t)) = i\left(\tilde{m}_n, \frac{\partial \tilde{m}_n}{\partial X}\right) \dot{X} \tag{9}$$

Denoting:

$$i\left(\tilde{m}_n, \frac{\partial \tilde{m}_n}{\partial X}\right) \equiv C_h \tag{10}$$

obtain:

$$C_h = \frac{\partial \Theta(X(t))}{\partial X} \tag{11}$$

Relation (10) is a definition of choice analogous to the definition of activity in (3). However, activity is a dynamical phase determined by the increment of $DM(\frac{\partial M}{\partial t})$, whereas choice is a topological phase determined by the total capital contribution of each financial channel via the change of $DM(\frac{\partial \widetilde{m}_n}{\partial X})$. In the former case, the increment of the entire $DM$ depends on time explicitly, while in the latter the change in each contribution $\widetilde{m}_n$ to M is associated with the contour (trajectory) of $\widetilde{m}_n$. If $\frac{\partial \widetilde{m}_n}{\partial X}$ is driven out of a given channel of money flow then $\cos\left(\widetilde{m}_n, \frac{\partial \widetilde{m}_n}{\partial X}\right) \to 0$ and choice tends to zero. It indicates that with unchanged $X_1$(prices), $X_2$(goods), $X_3$ should represent capital.

Relation (11) – the derivative of the topological phase – defines the potential of a gauge field, which in this context manifests itself as economic choice. The mechanism of market functioning is as follows. Emerging capital generates choice, while changes in choice define the fundamental forces – competition and profit:

$$\vec{C} = -\frac{\partial C_h}{\partial t}$$

$$\vec{P} = \text{rot } \vec{C}_h,$$

which organize market activity and maintain equilibrium through the emergence of expectations. From (11) we obtain

$$\Theta = \int_C \vec{C}_h d\vec{X},$$

where integration is carried out along the trajectory of $\widetilde{m}_n$. Two cases are possible. In the first, the trajectory is open and the phase $\Theta$ is equal to the difference of $\Theta(X(t_k))$ and $\Theta(X(t_0))$ between its values at the final and initial points of the trajectory. In the second, the trajectory is cyclic and the contour is closed. Then:

$$\Theta(t_k) = \oint_C C_h(\vec{X}) \, d\vec{X} \tag{12}$$

$\Theta$ – is the Berry phase [17] determined solely by the closed contour.

Using Stoke's theorem, (12) can be written as:

$$\Theta = \int_S \vec{P} \, d\vec{S} \tag{13}$$

where $\vec{P} = \text{rot } \vec{C}_h$ – profit, $S$— surface bounded by trajectory.

Equation (13) indicated that Berry phase constitutes the total profit of the market being the ultimate goal of the whole economics.

3. **Equations of Market Dynamics**

The first equation of dynamic market equilibrium with inflation in the dimensionless form is given by [2, 7]:

$$\frac{\partial C_\mu}{\partial X_\mu} = -\frac{\partial \ln M}{\partial t} \qquad (14)$$

where $C_h$ – competition flow component, $\frac{\partial \ln M}{\partial t}$ – flow of money supply growth (DM).

The economic meaning of (14) is the following: a deficit of DM generates competition and stimulates economic activity in the market. The growth of DM acts as a source of competition, while an increase in the competition flow hinders the further growth of DM.

The next equation:

$$\varepsilon_{\mu k} \frac{\partial C_k}{\partial X_v} = -\frac{\partial P_\mu}{\partial t} \qquad (15)$$

relates the curl (vorticity) of the competition flow to the evolution of the profit flow $P_\mu$. $\varepsilon_{\mu k}$ – Levi-Civita symbol. The profit flow is one of the phases of competition: when profit tends to increase, competition resists this growth; when profit tends to fall, competition works to raise it. Equation (15) is one of the laws governing the equilibrium of a market with arbitrarily high, but non-destructive, inflation.

Profit flow equation is given by $\vec{P}$:

$$\frac{\partial P_\mu}{\partial X_\mu} = 0 \qquad (16)$$

which expresses the absence of profit sources within the set of money flows and implies that the total change in profit in a closed market is zero. Last equilibrium equation:

$$\varepsilon_{\mu v k} \frac{\partial P_k}{\partial X_v} = \frac{\partial C_\mu}{\partial t} + K_\mu \qquad (17)$$

links key market characteristics: the internal circulation of profit (left-hand side of (17)), the competition flow, and capital $K_\mu = \frac{\partial \ln M}{\partial X_\mu}$ [7]. Each of the components of (17) corresponds to a particular mechanism supporting equilibrium, associated with the flows of capital, profit, competition, choice, and their dynamics. Collectively, these mechanisms constitute the local action of the invisible hand of the market. To visualize this mechanism, we must find the solutions of (14)-(17). Before solving them, however, it is useful to summarize how these equations arise.

Structural changes in market dynamics (inflation, etc.) radically modify the motion of DM flows, so that their straight-line motion in PS is replaced by parallel transport. Under parallel transport, moving source-flows of DM acquire an additional inhomogeneous phase — the Berry phase $\Phi$, which determines the total profit of the market $(\Theta = \Phi)$ in a way that

$$\frac{\partial \Phi}{\partial X_\mu} = C_h^{(\mu)}$$

defines the choice flow as the principal curvature of the trajectory of DM in phase space, and

$$\left(\frac{\partial^2 \Phi}{\partial X_\mu \partial X_\nu} - \frac{\partial^2 \Phi}{\partial X_\nu \partial X_\mu}\right) \neq 0$$

is the innate curvature of the surface swept out by this trajectory. These two quantities specify market dynamics under the new conditions. Curvature $\vec{C_h}$ curvature of the surface swept out by this trajectory. These two quantities specify market dynamics under the new conditions

$$\vec{C} = -\frac{\partial \vec{C_h}}{\partial t}, \vec{P} = \text{rot } \vec{C_h}$$

are the flows of competition and profit. Also, from the definition of $\vec{C}$ and equation (15) it follows that $\vec{C}$ is determined only up to the gradient of a particular function φ(X,t). We can therefore write $\varphi(X, t)$ and then:

$$C_\mu = -\frac{\partial C_h^{(\mu)}}{\partial t} - \frac{\partial \varphi}{\partial X_\mu} \tag{18}$$

The quantity φ represents the work of external forces associated with moving DM in the market and corresponds to the price of market products.

$$\frac{\partial \varphi}{\partial X_\mu} = \pi_\mu$$

So, $\pi_\mu$ is the price level, which defines the rate of inflation through the relative chang $\varphi(\pi = \frac{\varphi - \varphi_0}{\varphi_0} = \frac{\partial \varphi}{\partial X})$. Therefore, competition can be decomposed into price competition $(-\vec{\pi})$ and non-price competition $(-\frac{\partial \vec{C_h}}{\partial t})$, determined by the evolution of choice. The larger $(-\vec{\pi})$ is the more uniform prices are in the market. Non-price competition is generated by the evolution of choice. The higher non-price competition is, the more uniform choice becomes over the long run. The inhomogeneities of choice and price evolution are connected by

$$\frac{\partial C_h^{(\mu)}}{\partial X_\mu} + \frac{\partial \varphi}{\partial t} = 0 \tag{19}$$

which shows that changes in choice offset changes in prices. Thus, as choice increases (and since choice determines supply), supply grows and the price level falls. We now briefly discuss some consequences of (14), (18), and (19).

1) **Phillips Curve**

Combining (14), (18) and (19), we obtain

$$\text{div}\,\vec{\pi} = \frac{\partial \ln M}{\partial t} + \frac{\partial^2 \varphi}{\partial t^2} \qquad (20)$$

2) **Growth of price level** (inflation condition) is driven by the increment of DM and by the acceleration of price growth (the combined effect of monetary and accelerated price expectations).

By taking gradient of (14), we obtain

$$-\frac{\partial \vec{C_h}}{\partial t} = -\frac{1}{4\pi} \int \frac{\frac{d\vec{K}}{dt}\,d\vec{X'}}{|\vec{X} - \vec{X'}|} + \pi \qquad (21)$$

3) **Decline in choice flow** (non-price competition) is determined by capital shortage and the level of inflation; inflation may be described by relation (21)

$$\pi = \frac{1}{4\pi} \int \frac{\frac{d\vec{K}}{dt}\,d\vec{X'}}{|\vec{X} - \vec{X'}|} - \frac{\partial \vec{C_h}}{\partial t} \qquad (22)$$

4) **Accumulation of capital** and the decline of choice contribute to the rise of inflation $\pi$. Considering the connection between capital growth rate $\partial K/\partial t$ and unemployment rate u in the form [24]:

$$\frac{\partial K}{\partial t} = (u)^{-1}$$

Then as a result we obtain a dependence of inflation on the unemployment rate — the Phillips curve — in which a higher inflation rate is associated with falling unemployment and declining choice.

Denoting $\vec{\pi}$ as the cumulative change in the price level throughout the time period $(0, t)$, we can express the growth rate of this price level using formula (22).

Taking the gradient from (19), we obtain:

$$\vec{C_h} = -\frac{1}{4\pi} \int \frac{\frac{\partial \vec{\pi}}{\partial t'}}{|\vec{X} - \vec{X'}|} dX' \qquad (23)$$

5) A reduction in inflation expectations increases choice. Substituting (23) into (22) yields

$$\vec{\pi} = \frac{1}{4\pi} \int \frac{\left(\frac{\partial K}{\partial t'} + \frac{\partial^2 \pi}{\partial t^2}\right) dX'}{|\vec{X} - \vec{X'}|} \qquad (24)$$

6) Decrease in unemployment rate (movement of capital rate) together with accelerating inflation expectations leads to higher inflation.

The relations (20)-(24) give more complete description of the link between inflation and unemployment than the classical Phillips curve. In them, inflation depends not only on unemployment but also on choice, monetary and price dynamics and on inflation expectations – a dependence whose absence has often been cited as a weakness of Phillips' original theory [19]. In addition to the competitive force discussed above, there is a second vortical force – profit – which maintains market equilibrium and redirects money flows in directions that maximize profit. A more detailed discussion of (14)-(17) can be found in [2,3,7]. We now turn to the solution of these equations.

## 4. Solutions to dynamics equations

The relations (14)-(17) are nonlinear wave equations describing the spatio-temporal evolution of economic characteristics in local areas (different $\{X_i\}$) of the phase space. Consequently, the evolution of $\vec{C}_h, \vec{C}, \vec{P}$ and other related quantities is given by local nonlinear wave processes. The strong locality of market behavior in phase space is due to the fact that different combinations of prices, goods, capital, etc. – $\{X_i\}$ correspond to markedly different behaviors. Choice: $\vec{C}_h(\vec{X}, t)$, profit: $\vec{P}(\vec{X}, t)$ and competition $\vec{C}(\vec{X}, t)$ functions reflect this relationship. Problems of this type in mathematics, physics, biology and other fields are solved using methods from differential geometry and the theory of nonlinear equations [20]. One of the principal achievements of this approach is the discovery of close relations between the shape of principal curves (of the key quantities in a problem) and their physical, biological or other functional roles [21,22]. In our case, the principal quantities are choice $\vec{C}_h(\vec{X}, t)$ and its derivatives. It is therefore natural to expect that one can establish analogous relations in the economic context by finding expression for the choice function and related quantities.

We now proceed to describe functional specifications of variables $\vec{C}_h(\vec{X}, t), \vec{P}(\vec{X}, t), \vec{C}(\vec{X}, t)$, using methods covered previously [20, 22]. Since in each economic problem the variables ($X_1$, $X_2$, $X_3$, …) have different economic meaning, it's convenient to represent choice curves $\vec{C}_h(\vec{X}, t)$, in the parametric form $\vec{C}_h(s,t)$, where the main parameter is the length of the curve:

$$s = \int_0^s \sqrt{\sum_i \left(\frac{\partial X_i}{\partial s'}\right)^2} \, ds'$$

All calculations are held up in the local space. $\vec{t}$ – tangent, $\vec{n}$ – normal, $\vec{b}$ – binormal [23].

The final change in the shape of the choice curve $\vec{C}_h(\vec{X}, t)$ and related curves is obtained by transforming local relations into the global coordinate system as it's done in the string theory protein structure modelling [21], electro and hydrodynamics [22], plasticity [7]. Firstly, expression for $\vec{C}(\vec{X}, t)$ is obtained using equations (15), (17). In (15) we introduce vorticity of competition $\vec{\omega} = \text{rot } \vec{C}$, which is equal to the change in profit $\frac{\partial \vec{P}}{\partial t}$, and taking rot of (17), accounting for the fact that div $\vec{P} = 0$ and rot $\vec{K} = 0$, we obtain that $\Delta \vec{P} = \frac{\partial \vec{\omega}}{\partial t}$, or

$$\vec{P} = \frac{1}{4\pi} \int \frac{\vec{\omega}_t(\vec{X}')}{|\vec{X} - \vec{X}'|} d\vec{X}' \qquad (25)$$

Substituting (25) into (17) and evaluating the integral

$$\int \frac{\nabla_{\vec{X}} \times \vec{\omega}_t(\vec{X}')}{|\vec{X} - \vec{X}'|} d\vec{X}',$$

We find that if $\vec{X}, \vec{X}'$ don't explicitly depend on time:

$$\vec{C}(\vec{X}, t) = \frac{1}{4\pi} \int \frac{(\vec{X} - \vec{X}') \times \vec{\omega}(\vec{X}') d\vec{X}'}{|\vec{X} - \vec{X}'|^3} - K \qquad (26)$$

where $\vec{\omega} = \text{rot } \vec{C}$, and $\vec{K}$ – total capital at the given time moment $t$. We also assume that in a local region of economic phase space with coordinates $(X_1, X_2, X_3)$ the choice curve has length $2L$ and a certain small cross-sectional area.

After several calculations [22] for $\vec{C}$ in the local space we obtain:

$$\vec{C} = \frac{\gamma}{4\pi} \int \frac{(\vec{X} - \vec{X}') \times d\vec{e}}{|\vec{X} - \vec{X}'|^3} - K \qquad (27)$$

where

$$\gamma = \int_A |\vec{\omega}| \, dA = \int_e \vec{C} \, d\vec{e}$$

is the circulation of stable competition around the contour enclosing the cross-sectional area A. In its sense quantity $\gamma$ is the work done by the vortical competitive force, while $\frac{\partial \vec{C}_h}{\partial t}$ describes the displacement of DM increments along the contour determined by the sum of prices and quantities of goods, fixed for a given surface $S$ (along the circle of radius $r = \sqrt{x_1^2 + x_2^2}$). Taking integral of (22) and assuming $L \gg x$:

$$\vec{C}(s, t) = \frac{\gamma \, x(s, t)}{4\pi} \ln \frac{2L}{r} \vec{b}(s, t) - K \qquad (28)$$

where $x(s, t)$ – is the curvature of the choice curve segment, $r$ is the radius of the cross-section, and $\vec{b}(s, t)$ is the binormal. All quantities depend on the parameter $(s)$ and time $(t)$.

The resulting non-price competition is directed along with the binormal and is proportional to the curvature of the curve $(s, t)$ rising with the increase in $L$ and falling with the growth of $r$ according to the logarithmic law.

We now estimate the effect of capital $K$ on the competitive system, i.e. on the transverse boundary at $r = d$, beyond which $\vec{C} = 0$:

$$K = \frac{\gamma\, x(s,t)}{4\pi} \ln\frac{2L}{d} \vec{b}(s,t)$$

Then

$$\vec{C}(s,t) = \frac{\gamma\, x(s,t)}{4\pi} \ln\frac{d}{r} \vec{b}(s,t) \tag{29}$$

That is, capital reduces non-price competition, effectively trading off the length of the choice region against its transverse dimensions. Then we consider the dynamics of the choice curve in the local coordinate system, using the equation

$$\frac{\partial C_h(s,t)}{\partial t} = x(s,t)\, \vec{b}(s,t) \tag{30}$$

Parametric equation of the curve [23] is given by:

$$\frac{\partial \overrightarrow{C_h}(s,t)}{\partial s} = \vec{t} \tag{31}$$

And the Frenet equations [23]

$$\begin{aligned}
\frac{\partial \vec{t}}{\partial s} &= x\,\vec{n}, \\
\frac{\partial \vec{n}}{\partial s} &= \tau\,\vec{b} - x\,\vec{t} \\
\frac{\partial \vec{b}}{\partial s} &= -\tau\,\vec{n}
\end{aligned} \tag{32}$$

where $t$ – market, $\tau$ – torsion of the choice flow.

In this formulation, the shape of the curve is determined by [23]:

$$\psi(s,t) = x(s,t)\, \exp\left\{ i \int_0^s \tau(s',t)\, ds' \right\} \tag{33}$$

which satisfies a nonlinear Schrödinger equation and represents its solution — a helical curve with constant torsion $\tau$ and curvature x(s,t) [22, 23]:

$$x(s,t) = 4\beta\, \text{sech}\, 2\beta(s - 2\tau t) \tag{34}$$

which changes based on the maximum value equal to $4\beta$ at a point $s = 2\tau t$, to zero as $s \to \infty$. This shifted curve is the axis of the choice region with a segment of enhanced vorticity of competition (profit increment) of order $\frac{4}{2\beta}$. It moves along a vortical parametric trajectory with velocity $(-2\tau)$. Both curvature and torsion are scalars and are determined by the particular shape of the choice curve. The change in the shape x(s,t) is driven by a localized nonlinear wave – a topological soliton that sets the rate of change of choice i.e. of non-price competition (with the opposite sign). The soliton wave propagates freely along the binormal $\vec{b}$ and along the curve. Since $\vec{b}(s,t)$ is oriented and torsion $\tau$ is non-zero, the polarization of the wave undergoes rotation in economic phase space by an angle

$$\Theta = \int_0^L \tau(s)\, ds,$$

equal to the solid angle swept out by the tangent vector $\vec{t}$ on the unit sphere [23]. In the special case of constant torsion this gives $\Theta = \tau L$. This is the law of parallel transport of the vector $\vec{C}$ in the economic phase space with the presence of inflation.

The choice curve is analogous to a string [21], for which curvature x and torsion $\tau$ can be regarded as gauge transformations of the form

$$x \sim e^{i\Theta(s)} x, \tau \sim \tau + \partial \Theta(x)$$

In this interpretation, curvature plays the role of a Higgs field, while torsion is a gauge field [21]. The invisible hand of the market thus acts as an Abelian Higgs-type mechanism [21]. This action is indeed difficult to visualize, but it is possible to make visible key market characteristics – $\vec{C_h}$, $\vec{C}, \vec{P}$. One can express $\vec{C_h}, \vec{P}$ similarly to $\vec{C}$ in the local coordinates but it is more informative to express them in the global coordinates $\{x_i\}$, thereby reconstructing the overall picture of how the «visible hand» of the market operates.

The shape of the curve in the general coordinate system $(x_1, x_2, x_3)$ for given curvature and torsion is determined via standard differential-geometric constructions and solutions of Riccati-type equations [21,22]. The resulting components $C_h^{(1)}, C_h^{(2)}, C_h^{(3)}$ are given in the following form:

$$C_h^{(1)} = \frac{1}{\beta(1+v^2)} \left( \frac{\sin vx}{\operatorname{ch} x} + \frac{\sin vy}{\operatorname{ch} y} \right) \tag{30.1}$$

$$C_h^{(2)} = \frac{l}{\beta(1+v^2)} \left( \frac{\cos vx}{\operatorname{ch} x} - \frac{\cos vy}{\operatorname{ch} y} \right) \tag{30.2}$$

$$C_h^{(3)} = s - \frac{l}{\beta(1+v^2)} (\operatorname{th} x + \operatorname{th} y) \tag{30.3}$$

where $x = 2\beta(s - 2\tau t), y = 4\beta\tau t, v = \frac{2\tau}{\beta}$.

Two choice components $C_h^{(1)}, C_h^{(2)}$ corresponding to the price and goods $C_h^{(P)}, C_h^{(q)}$ choice may exhibit oscillatory behavior. Component $C_h^{(3)}$ lies along the capital axis and shows no such oscillations. The dimensionalities of the quantities entering (29)/(30) are

$$[\beta] = \left[\frac{1}{s}\right], [\tau] = \left[\frac{1}{s}\right], [t] = [\gamma] = [s^2]$$

To construct and analyze $\vec{C_h}(s,t)$ we choose specific values with $\beta = 2\tau$ in price units. Figures 1-4 show the components $C_h^{(1)}, C_h^{(2)}, C_h^{(3)}$ and $C_h = ((C_h^{(1)})^2 + (C_h^{(2)})^2)^{\frac{1}{2}}$ if $\beta = 0.5, L = 5$. Expression $((C_h^{(1)})^2 + (C_h^{(2)})^2)^{1/2}$ – shows the value of the choice in the plane $(P, Q)$. Components $C_h^{(1)}$ and $C_h^{(2)}$ representing price and quantitative choice behave differently in response to changes in $s, t$. That is, $C_h^{(1)}$ fall if $s, t$ increase, while $C_h^{(2)}$ grows in this case. Both exhibit temporal oscillations. The capital-driven choice $C_h^{(3)}$ — does not oscillate; it grows monotonically with s and only later changes with t for each fixed s. Its magnitude $\left|C_h^{(3)}\right|$ is greater than the magnitudes $\left|C_h^{(1)}\right|, \left|C_h^{(2)}\right|$ and in fact this factor is the most influential component of the total choice magnitude $\left[\sum_i (C_h^{(i)})^2\right]^{1/2}$

**Figure 1. Price choice component $C_h^{(1)}$, $\beta = 1/2$**

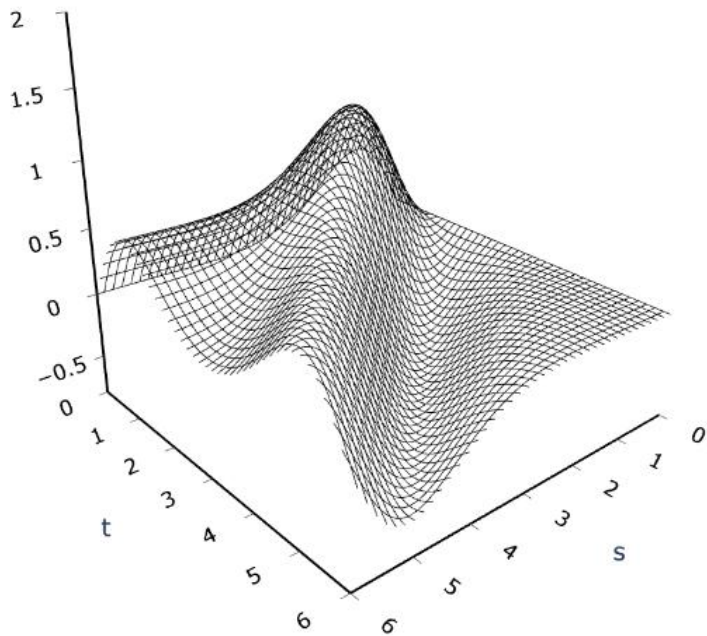

**Figure 2. Quantitative choice component $C_h^{(2)}$, $\beta = 1/2$**

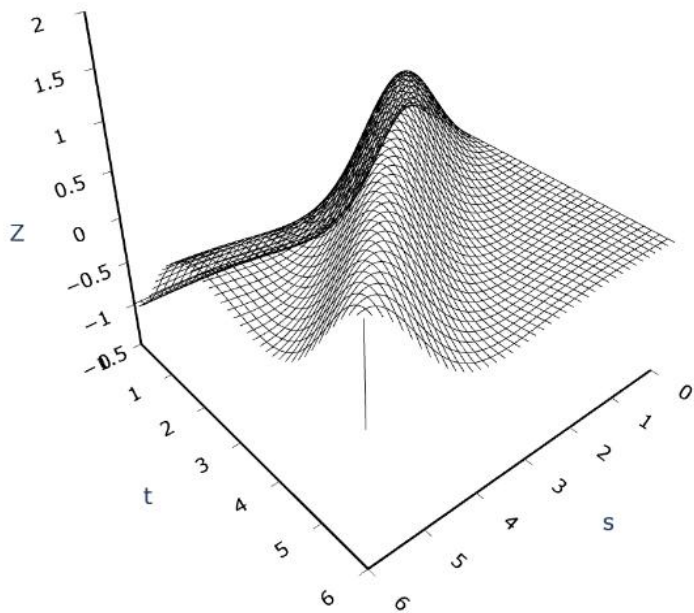

**Figure 3. Capital-driven choice $C_h^{(3)}$, $\beta = 1/2$**

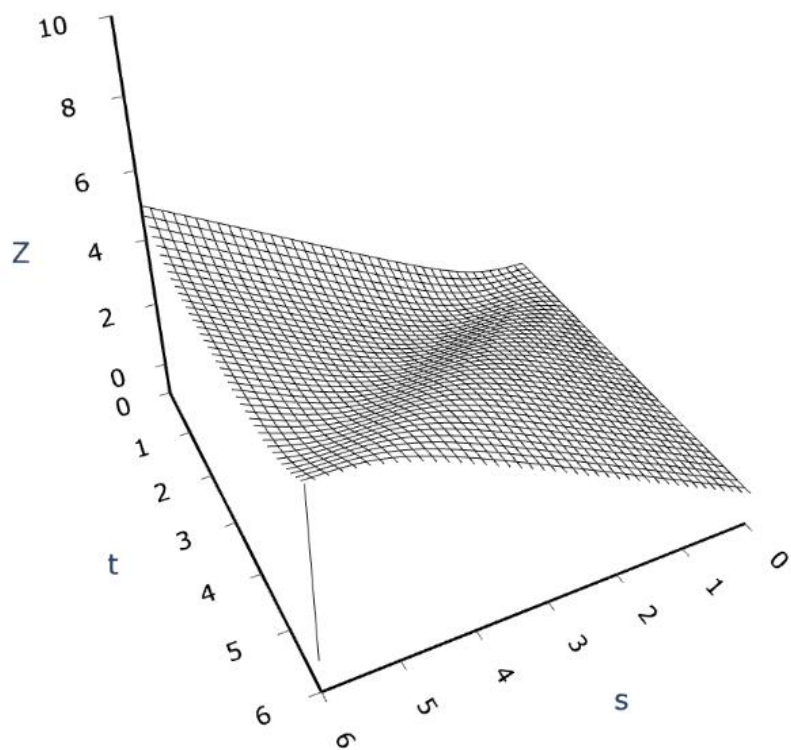

**Figure 4. Choice value $|\overrightarrow{C_h}|$, $\beta = 1/2$**

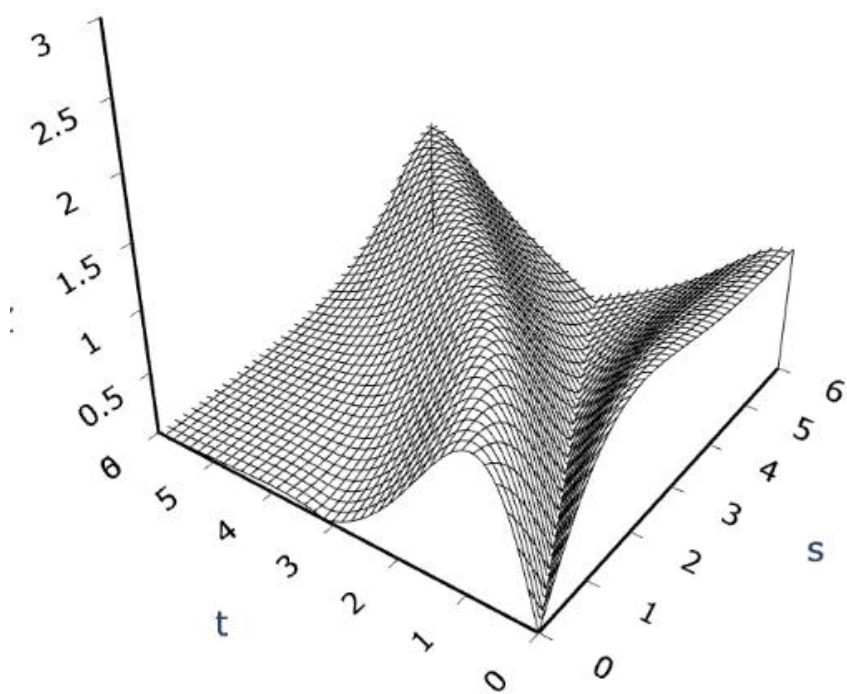

A particularly notable feature of choice is shown in Figure 4 which illustrates the choice surface in the $(P, Q)$ plane. Two maxima appear at the onset of market interactions $(s, t \to 0)$. One of them is present for all $s$ and if $t \to 0$ while the other drifts toward larger $s$ over time. These maxima are separated by a region that expands with increasing $s$. The magnitude of choice is determined mainly by $C_h^{(3)}$, driven by the component $X_3$ (capital) and $P, Q$, indicating the market supply, have only a weak influence on $\vec{C_h}$. Consequently, choice (demand) determines supply, and that the shape of the choice curve governs market behavior.

Furthermore, Figures 1–4 show that choice is non-cyclic and dual in nature: on the one hand, it is an individual action; on the other, it is an economic category that maintains equilibrium under inflation. In this hierarchy equilibrium takes precedence over the satisfaction of needs. Meeting needs allows rational choice to dominate but even non-rational choice can sustain equilibrium. Thus, it is not prices and goods that act as the invisible hand of the market, but choice and its derivatives. They restore equilibrium disturbed by inflation. The magnitude of choice is determined mainly by capital choice, while price and quantity choice shape its oscillatory features.

Figure 5 is the most illustrative. $\left[(C_h^{(1)})^2 + (C_h^{(2)})^2\right]$ based on $(p, q)$ where $p = \pm\frac{\Delta p}{p}$, $q = \pm\frac{\Delta q}{q}$. The choice interacts with market participants as follows. The characteristic duration of a choice event is much shorter than the characteristic time of market behavior. So, the market is forced to absorb and emit competition $\vec{C}$. If absorbed and emitted amounts balance, the market returns to its initial state after the passage of the choice wave – becoming transparent to choice.

$\vec{C_h}$ and $\vec{C}$ are nonlinear stationary wave processes (as visualized above) with different similar $s$ and slightly different propagation speeds. They eventually "catch up" with one another and cause their trajectories to intersect. Upon collision, the waves exchange activity, and their like-signed solutions transform into a pair of solitons. The second soliton appears in the region $\Delta p/p < 0$, $\Delta q/q < 0$. This exchange of activity occurs away from the main body of the market. With sufficiently strong competition, the first soliton pushes the market into an excited state, transferring activity to it. Under the action of the second soliton, the ensemble of agents returns activity to it. The net result is that choice remains unchanged due to this activity exchange.

When the states of choice $\left(\frac{\Delta p}{p}, \frac{\Delta q}{q}, \frac{\Delta K}{K}\right)$ and the states of market agents approach one another, agents absorb activity and transition from non-buyers to buyers. The leading soliton weakens. The trailing soliton interacts with agents and returns them to a non-buyer state. This process emits activity, restoring the choice wave to its equilibrium soliton form and allowing it to propagate at constant speed. Thus, choice transforms the market into a medium transparent to itself.

This phenomenon is analogous to self-induced transparency in electrodynamics (optics). Therefore, the "hand of the market" is not a metaphor or a theoretical abstraction, but a real mechanism analogous to self-induced transparency in nature. To compute and visualize $\vec{C}, \vec{P}$ we use a connection between Berry phase and choice. Integrating (11) we obtain:

$$\mathbb{P}^{(3)} = \Theta^{(3)} = \int C_h^{(3)} dX = \frac{1}{\beta(1+v^2)} \ln \frac{\operatorname{ch} X}{\operatorname{ch} Y} \tag{31}$$

$$C^{(3)} = -\frac{\partial C_h^{(3)}}{\partial t} = \frac{\beta A}{(1+v^2)} (\operatorname{sech}^2 Y - \operatorname{sech}^2 X) \tag{32}$$

$$P^{(3)} = \varepsilon_{312} \partial_1 C_h^{(2)} = \frac{2\beta\tau}{(1+v^2)} * \frac{Xt}{r^2} \operatorname{sech}^2 Y \tag{33}$$

where $X = 2\beta(S + \frac{\tau r}{4\pi L} t)$, $Y = \frac{\beta\tau t}{2\pi L}$, $r^2 = X_1^2 + X_2^2$.

$\varepsilon_{312}$ – Levi-Civita symbol. The dependence on $P, Q$ is stated explicitly.

**Figure 5. Two-soliton choice form**

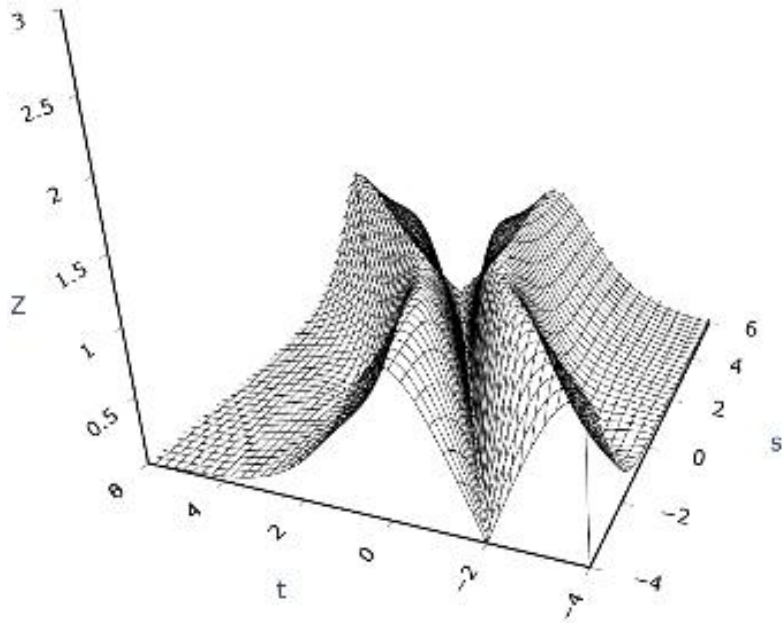

**Figure 6. Component $C^{(3)}$ of non-price competition**

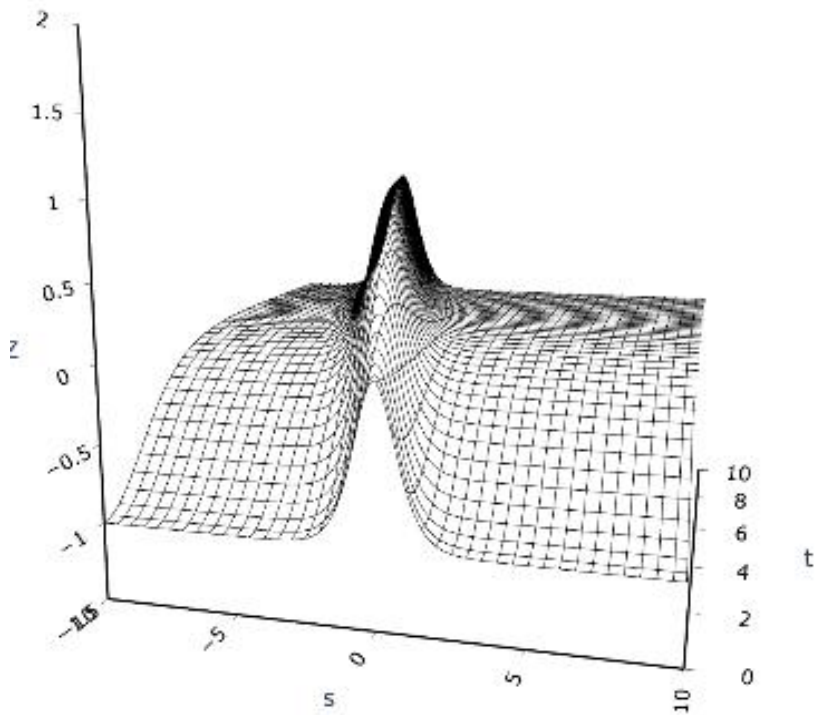

**Figure 7. Component $\mathbb{P}^{(3)}$ of the total profit**

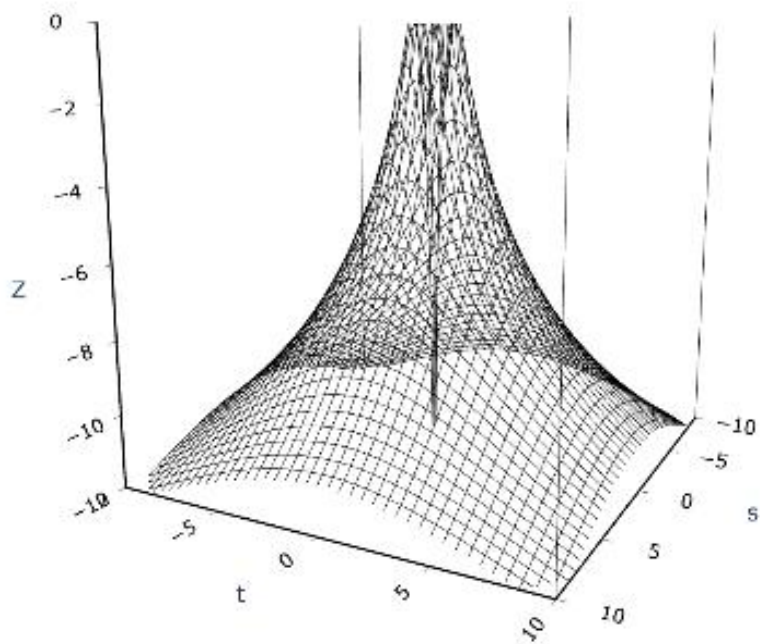

### Figure 8. Component $P^{(3)}$ of the profit

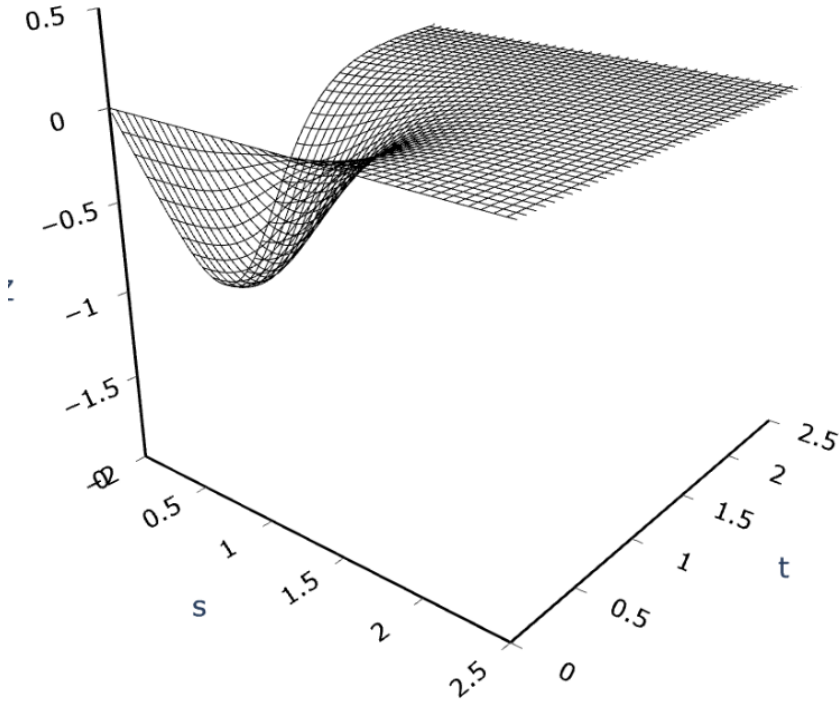

Figures 5–8 complement Figures 1–4 and reveal the mechanism through which inflationary market transitions from one equilibrium to another. Before inflation the money mass moved along continuous trajectories in economic phase space $(X_1, X_2, X_3)$. In dimensionless variables $\frac{\Delta P}{P}, \frac{\Delta Q}{Q}, \frac{\Delta K}{K}$ – representing relative changes in price, quantity, and capital – an increment of DM only modified market activity

$A \sim (\vec{M}, \frac{\partial \vec{M}}{\partial t})$, defined by the dynamic phase function $(\frac{\partial \vec{M}}{\partial t} / \vec{M},)$.

Under deficit DM and inflation previous equilibrium breaks down as trajectories of DM become interrupted by "defects"—regions lacking DM. To maintain motion under these new conditions, phase space must become curved (fibered), and motion switches to parallel transport. In this geometry emerges a compensating field

$$C_h \sim (\vec{M}, \frac{\partial \vec{M,}}{\partial X}),$$

associated with a Berry-type topological phase $\Theta(X)$. This quantity

$$C_h \sim \frac{\partial \Theta(X)}{\partial X}$$

is determined by capital $(\frac{\partial \vec{M}}{\partial X})$ and represents choice which forces the trajectory to bypass the emerging defects. Two forces created by this field assist in this process: the evolution of choice –

$\frac{\partial C_h}{\partial t}$ – which corresponds to non-price competition, and the vorticity of choice that is profit – $\text{rot}\vec{C_h}$.
In addition, the equations describing the mutual dependence of $\vec{C_h}, \vec{C}, \vec{P}$ allow for price competition in the market itself, associated with the formation of internal prices. This latter quantity also determines the inflation index being external with respect to the "invisible hand of the market". The role of both types of competition is to create conditions for resonant interaction of agents with choice. That is, competition drives the economic states of agents toward states close to the state of choice (demand). As long as this correspondence is not achieved, nothing significant happens on the market. In essence, we are dealing with the reconciliation of demand and supply, i.e. the region where these two categories are in equilibrium. Hence, the principal "unpleasantness" of inflation is the deterioration of the economic condition of agents and the impossibility of conducting normal market transactions.

It should be emphasized that equilibrium for each economic state $(P, Q, K)$ does not lie on a single and universal choice curve. Each state $(P, Q, K)$ has its own equilibrium point on the choice surface. The choice (demand) curve in $(P, Q)$ is thus the locus of all such equilibrium points. In this light, it is at least perplexing that the so-called law of demand $(P \sim \frac{1}{Q})$ is often interpreted as describing a single curve for different economic states, whereas, strictly speaking, each state has its own equilibrium on its own choice contour.

That is, for each price P in a given market we actually have a different choice curve and a different equilibrium. So, the «Great Law of Demand» becomes just an illusion.

As an illustration, consider $|C_h^{(P,Q)}|$ if: $L = 2\pi$, $s = \pi$, $0 \leq P, Q \leq 1.5$, $t = 1$.

In this case

$$|C_h^{(P,Q)}| \simeq \sec(\frac{(P^2 + Q^2)^{1/2}}{a}), a = 16\pi^2.$$

Hence:

$$P^2 + Q^2 = (a \cdot \text{arcsech}\,(|C_h^{(P,Q)}|))^2,$$

or

$$P^2 + Q^2 = R^2(P, Q),$$

which is the equation of a circle of radius $R = R(P, Q)$. thus, for any change in $P, Q$ the value of $R$ and hence the curve $P(Q)$ will be different for each combination $(P, Q)$. In facts the larger value of $P$ lead to the smaller radius $R$. The law of demand in this framework can be expressed in the form:

$$\boxed{f(P, Q) = \mathcal{F}^{-1}(|C_h^{(P,Q)}|)}$$

where $f(P, Q)$ – functional dependence $P(Q)$, along the demand curve, determined by the magnitude of price-related choice.

**Conclusion**

In this work we showed that economics in its sense is very similar to natural sciences, including physics. We outlined that often unclear influence of the "invisible hand" in economic markets can be made "visible" by applying the methods used in gauge filed theory and plasticity. In fact, light (as an electromagnetic field), Darwinian natural selection, plastic deformation, and choice in economics - all evolutionary and adaptive processes operate according to the same underlying mechanism.

We have shown that structural changes in the markets break the initial equilibrium of money flows and lead to changes of the curvature in economic phase space. The temporal and spatial derivatives of this choice field generate the observable forces of competition and profit, governing market adjustment and equilibrium restoration. Total profit, being one of the key driving mechanisms of the economy, can be described using equations similar to the Berry phase formulas. Explicit solutions demonstrate that market adaptation proceeds through localized nonlinear waves of choice, analogous to self-induced transparency in electrodynamics, allowing the market to deal with competitive pressures. Taken together, these results provide a mathematically grounded interpretation of the "invisible hand" as a real, dynamically operating field mechanism in which choice, competition and profit are the key driving forces.